\def\year{2015}
\documentclass[letterpaper]{article}
\usepackage{aaai}
\usepackage{times}
\usepackage{helvet}
\usepackage{courier}
\usepackage{amssymb}
\usepackage{amsmath}
\usepackage{subfigure}
\usepackage[english]{babel}
\usepackage{graphicx}
\usepackage{array}
\usepackage{booktabs}
\usepackage{upquote}

\begin{document}
%
%
\title{Leveraging Social Foci for Information Seeking in Social Media}
\author{
%
%
Suhas Ranganath \\ srangan8@asu.edu \\Arizona State University \And Jiliang Tang\\Jiliang.Tang@asu.edu\\Arizona State University  \And  Xia Hu \\Xia.Hu@asu.edu\\Arizona State University  \AND Hari Sundaram \\hs1@illinois.edu\\ University of Illinois Urbana Champaign \And Huan Liu\\ Huan.Liu@asu.edu\\Arizona State University\\
}

\maketitle
\begin{abstract}
\begin{quote}
  The rise of social media provides a great opportunity for people to reach out to their social connections to satisfy their information needs. However, generic social media platforms are not explicitly designed to assist information seeking of users. In this paper, we propose a novel framework to identify the social connections of a user able to satisfy his information needs. The information need of a social media user is subjective and personal, and we investigate the utility of his social context to identify people able to satisfy it. We present questions users post on Twitter as instances of information seeking activities in social media. We infer soft community memberships of the asker and his social connections by integrating network and content information. Drawing concepts from the social foci theory, we identify answerers who share communities with the asker w.r.t. the question. Our experiments demonstrate that the framework is effective in identifying answerers to social media questions.
\end{quote}
\end{abstract}

\noindent Information seeking is defined as ``A conscious effort to acquire information in response to a need or gap in knowledge'' \cite{case2012looking}. Online social media makes it easier for users to reach out to a large number of friends, leading people to use them to seek information from their social connections. This gives rise to a distinct way for online information seeking, wherein the information needs expressed are subjective and personal to the asker. An interesting way people leverage online social media to seek information is by asking questions through their status messages \cite{morris2010people}. This phenomenon is prevalent in social media platforms like Twitter and Facebook and has received considerable attention in recent literature \cite{efron2010questions,paul2011twitter,Lampe:2014}.

However, unlike dedicated Q\&A platforms, generic social media sites like Twitter and Facebook are not designed for information seeking \cite{paul2011twitter}. Questions are not archived, thus finding people who answered similar questions in the past is difficult. Questions are buried among other content produced by the social connections of a potential answerer. Designing algorithmic frameworks to identify answerers to social media questions will help to bridge the information gap of users and increase user satisfaction. This framework can also help enhance Twitter search by making it personalized to the asker.

Information need of social media user is subjective or personal, unlike traditional Q\&A platforms like Stackoverflow, and hi social context is useful to find appropriate people able to satisfy it \cite{hecht2012searchbuddies}. Also, users with higher tie strength with the asker were shown to better satisfy information needs in social media \cite{panovich2012tie}. For example, to assist a person looking to get a new hairstyle, finding people from his social connections who share related context with him can be more useful to him than finding web pages related to hair salons.

This task faces several challenges. The questions are textual while the social context of the asker can involve network information. Integrating such kind of heterogeneous information will help to efficiently utilize social context to identify answerers to social media questions. Each social media user has many social connections and produces a lot of content leading to significant issues of scalability. Finally, the social context of the asker related to the question needs to be determined and appropriately utilized in order to identify suitable answerers.

\begin{figure}
\begin{center}
\subfigure[]{\label{fig:socfoc}\includegraphics[scale=0.25]{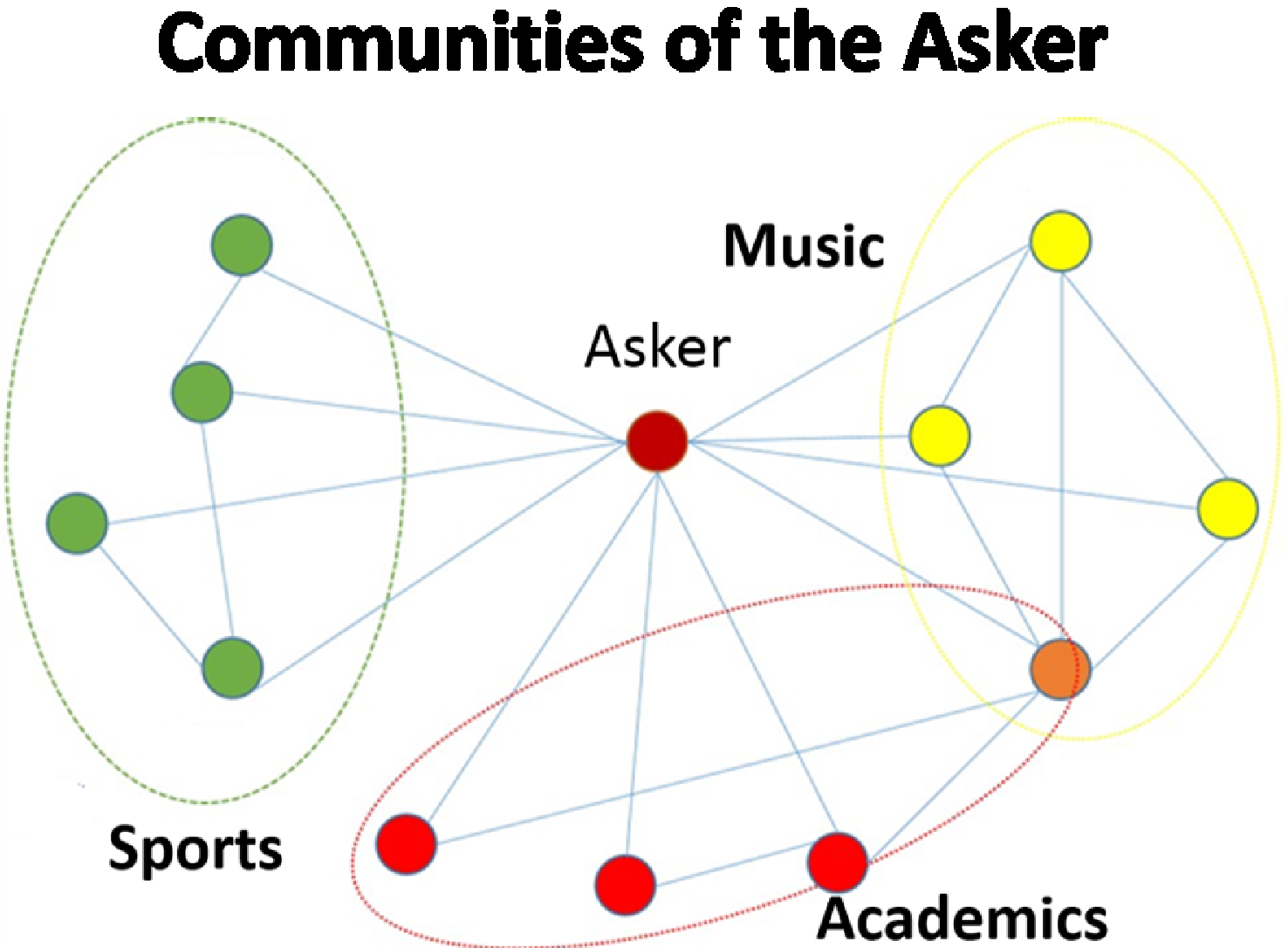}}
\hspace*{0.01in}\subfigure[]{\label{fig:socques}\includegraphics[scale=0.25]{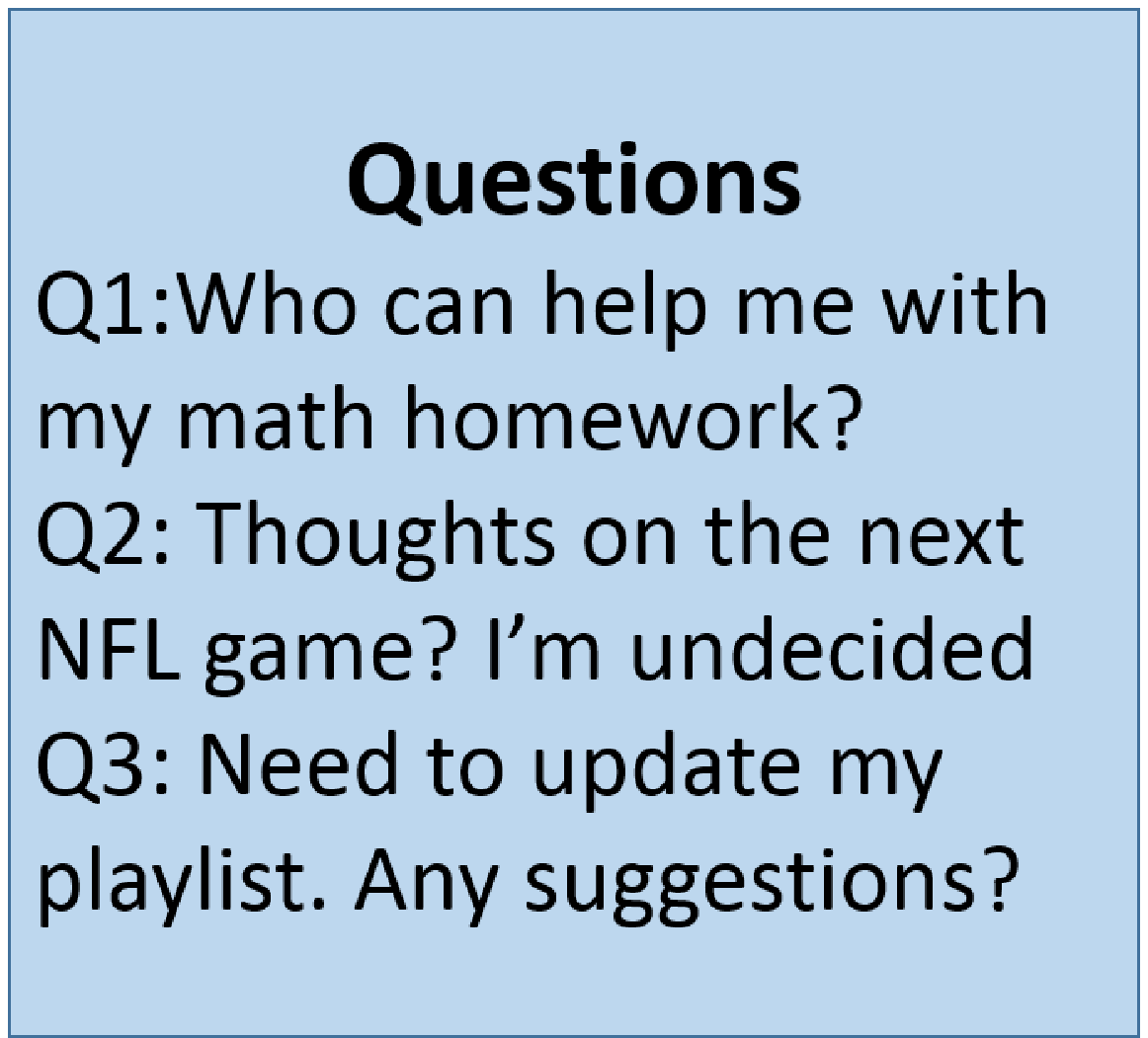}}
\vspace*{-0.2in}
\end{center}
\caption{(a) Different foci a user shares with his social connections. (b) Questions of users. Users sharing different foci with the asker are more likely to answer related questions.}
\label{fig:socfoctheo}
\vspace*{-0.2in}
\end{figure}

In sociological literature, the social foci theory postulates that interactions between people are organized around relevant entities known as foci \cite{feld1981focused}. A focus can be the activities, interests, and various affiliations of a user. Different groups of social connections of a user share different foci with him. For example, from Fig. \ref{fig:socfoc} we see that the user shares an interest of sports with his connections in green, an interest of music with his connections in yellow and academic interests with his connections in red.

Inspired by the social foci theory we propose that,  people in social media sharing social foci related to the question with the asker  are suitable to answer them. Illustrative examples of questions are given in Fig. \ref{fig:socques}. The asker of Q1 is seeking assistance in his math homework, and this might be best responded by users sharing academic foci with him. Q2 is seeking opinions on an NFL game, and this might be best provided by his connections sharing foci related to sports with the asker. Similarly, Q3 might be best answered by connections sharing music related foci with the asker.

In this paper, we propose a framework to investigate the utility of social context derived from network and content information in identifying answerers to social media questions. Informed by the concepts of social foci theory illustrated in Fig. \ref{fig:socfoctheo}, we utilize social context of an asker related to the question, and demonstrate that the framework is effective in identifying answerers for social media questions. Specifically, we address the following questions: How to utilize the network and content information of the asker and his social connections to better identify answerers for social media questions? Are approaches based on shared context in the question domain useful in identifying answerers to different kinds of social media questions?

The main contributions of our work are as follows:
\begin{itemize}
\item Formally defining the problem of finding suitable users to answer questions in online social media platforms,
\item Proposing a framework to exploit network and content information to identify answerers to questions, and
\item Conducting experimental evaluations of the framework on a dataset of social media questions.
\end{itemize}

\section{Related Work}
\label{sec:relworks}

Social media questions have received considerable attention in research communities \cite{yang2011culture,MEET:MEET14504701208,Lee:2012:UMQ:2207676.2208741}. An analytical study on questions asked and the answers received in Twitter is presented in \cite{morris2010people,paul2011twitter}. They indicated that subjective questions were the most prevalent and the trust users have on their friends was the primary factor for asking questions. A study of questions and responses received in Facebook was conducted in \cite{gray2013wants,ellison2013calling} and bridging social capital was proposed to be a strong motivation for Q\&A activity in social media. These works give interesting insights to the question answering process in social media, but do not focus on identifying answerers to these questions.

Systems to identify answerers for social media questions adopt different methods such as matching question content with profile information \cite{hecht2012searchbuddies} and using crowdsourced technology \cite{jeong2013crowd}. Social search architectures and empirical models to route questions to answerers using different kinds of social information are discussed in \cite{horowitz2010anatomy,nandi2013little}. These works are meant to demonstrate social search systems and hence do not contain any experimental evaluations.

A related line of research is the study of community Q\&A systems like Yahoo! Answers \cite{adamic2008knowledge} and Quora \cite{wang2013wisdom}. Content from existing Q\&A sessions are used to rank answerers by NLP techniques. \cite{jurczyk2007discovering} uses link structure to find authoritative answerers for a question category. \cite{zhou2012topic} and \cite{yang2013cqarank} combine network and content information to identify authoritative users as answerers. The environment for social media questions is different as the candidate answerers are themselves connected via social relations. Systems utilizing question categories \cite{zhu2013ranking} cannot be applied as they are not explicitly known in generic social media. Social expertise systems  \cite{pal2011identifying,bozzon2013choosing} identify subject matter experts in social media. Social media questions  are subjective and personal might require answerers who share social context with the asker rather than subject matter experts.

Another related field to our work is the application of social foci theory in social media. Social foci theory has received attention in several domains such as relational learning \cite{tang2009relational} and structural hole theory \cite{burt2009structural}. Recently, social foci theory has been used to derive community memberships using both node and edge attributes \cite{yang2014community}. To the best of our knowledge, this is the first work that has utilized concepts from social foci theory to identify answerers for social media questions.

\section{Problem Definition}
\label{sec:prosta}

We first describe some general notations. Boldface uppercase letters (e.g $\mathbf{X}$) denote matrices and boldface lowercase letters (e.g. $\mathbf{x}$) denote vectors. $\mathbf{X}_{ij}$ signifies the element in the $\text{i}^{\text{th}}$ row and $\text{j}^{\text{th}}$ column of matrix $\mathbf{X}$ and the $\text{i}^{\text{th}}$ row of a matrix $\mathbf{X}$ is denoted by $\mathbf{X}_i$. Similarly $\mathbf{x} _i$ denotes the $\text{i}^{\text{th}}$ element of vector $\mathbf{x}$ and $\mathbf{x}_\text{y}$ denotes a vector $\mathbf{x}$ corresponding to the quantity y. We denote the Frebonius norm of a matrix $\mathbf{X}$ as $||\mathbf{X}||_{F}=\sqrt{\sum_{i,j}{\mathbf{X}_{ij}^2}}$.

We now define some terms related to the questions asked, the network and content of the asker and his social connections. We define attributes of a question q as the set of words used in the question i.e. $\mathbf{w}_\text{q}=[\text{w}_{\text{q1}},\text{w}_{\text{q2}},...,\text{w}_{\text{ql}}]$. Since we are dealing with subjective questions, the asker marking the answer to be useful or publicly acknowledging the answerer gives the evidence of its acceptance.

Let A denote the asker of the question q and $\mathbf{f}_\text{A} = [\text{f}_{1},\text{f}_{2},....,\text{f}_{\text{m}}]$ denote the social connections of A and m is the number of social connections of A.  We define the egonetwork of each asker A as consisting of the asker, the social connections of the asker and the links among his social connections. The egonetwork of asker A, $\mathbf{N} \in \mathbb{R}^{(m+1)\times(m+1)}$ is given by
 \small\begin{equation*}
 \mathbf{N}_{ij} =
  \begin{cases}
   1 & \text{directed edge from f}_{\text{j}} \text{ to f}_{\text{i}},\text{i}\ne \text{j, i,j} \in {\{\text{A},\mathbf{\text{f}}_{\text{A}}\}}\\
   0 & \text{otherwise}
  \end{cases}
  \end{equation*}\normalsize
We collect the status messages of the asker and his social connections. We apply basic preprocessing steps such as removal of stop words and stemming. We then define the user-word matrix $\mathbf{S} \in \mathbb{R}^{(m+1)\times w}$ of asker A as
\small\begin{equation*}
 \mathbf{S}_{ij} =
  \begin{cases}
   \text{num*tfidf}_{\text{j}} & \text{if user } \text{u}_{\text{i}} \text{ has used word } \text{w}_{\text{j}} \text{ num times} \\
   0 & \text{if user u}_{\text{i}} \text{ has not used the word w}_{\text{j}},
  \end{cases}
  \end{equation*}\normalsize
where num is the number of times the user $\text{u}_\text{i}$ has used the word $\text{w}_\text{j}$, w is the total number of words used by the asker and his social connections and $\text{tfidf}_{\text{j}}$ is the tf-idf score of word $\text{w}_{\text j}$. A single user will only use a small subset of the total number of words, resulting in $\mathbf{S}$ being sparse.

With the terminologies and the notations described above, we formally define the problem as follows \textit{``Given a question q, an asker A, the network neighborhood of the asker $\mathbf{f}_{A}$, find a suitable set of people among $\mathbf{f}_{A}$ whose responses for the question q that the asker accepts''}.

\section{Information Seeking via Social Foci}
\label{sec:methodology}
In this section, we describe our framework to identify answerers for social media questions in detail. First, we infer social foci memberships of the asker and his social connections from their network and content information. We then compute the overlap in foci memberships of the asker and his social connections in the question domain to identify answerers to these questions.

\subsection{Modeling Content Information}
We model the content information to infer major foci of the asker and his social connections. We draw from Non-negative Matrix Factorization (NMF) presented in \cite{seung2001algorithms} to infer foci from the user-word matrix $\mathbf{S} \in \mathbb{R}^{(m+1)\times w}$. We factorize the matrix $\mathbf{S}$ into two low dimensional sparse non-negative matrices, $\mathbf{U} \in \mathbb{R}^{(m+1)\times k}$  and $\mathbf{P} \in \mathbb{R}^{w\times k}$ such that $\text{k} \ll \text{m}$ by solving the following optimization problem.
\small\begin{equation}
\min_{\mathbf{U}\geq 0, \mathbf{P}\geq 0} ||\mathbf{S-UP^{\text{T}}}||_{F}^{2}
\label{eqn:wordmat}
\end{equation}\normalsize
Here, k is the number of latent foci in the neighborhood of the asker and m is the number of his social connections. $\mathbf{U}$ denotes the latent foci membership of the asker and his social connections and $\mathbf{P}$ denotes the latent foci memberships of words. The correlation between foci memberships of the words can be obtained by the overlap in the corresponding rows of $\mathbf{P}$. The constraints $\mathbf{U}\geq 0$ and $\mathbf{P}\geq 0$ denote that the matrices have all non-negative elements. The non-negativity ensures an intuitive decomposition of the matrix into its constituent parts.

\subsection{Integrating Network Information}
In a social setting, the interests or affiliations of an user are correlated with the interests of his social connections, thereby affecting his memberships to different foci \cite{feld1981focused}. This notion is also supported by network homogeneity \cite{marsden1988homogeneity}, which says that people connected to each other display similar interests and affiliations. Therefore, it is essential to utilize network structure to determine foci memberships of the asker and his social connections.

To utilize the network structure, we first factorize the ego network of the asker $\mathbf{N}$ into two low rank non-negative matrices $\mathbf{U}  \in \mathbb{R}^{(m+1) \times k}$  and $\mathbf{V} \in \mathbb{R}^{k \times k}$ s.t. $\text{k} \ll \text{m}$ by solving the following optimization problem.
\small\begin{equation}
\min_{\mathbf{U}\geq 0,\mathbf{V}\geq 0} ||\mathbf{N-UVU^{\text{T}}}||_{F}^{2},
\label{eqn:netmat}
\end{equation}\normalsize
where $\mathbf{U}$ contains the membership of the asker and his social connections to different latent foci and $\mathbf{V}$ contains the correlations between the foci. The constraints $\mathbf{U}\geq 0$ and $\mathbf{V}\geq 0$ denote that the matrices have only non-negative elements.

We then integrate network and content information to infer the foci membership of the asker and his social connections by formulating the following optimization problem.
\small\begin{align}
\min_{\mathbf{U}\geq 0,\mathbf{V} \geq 0,\mathbf{P}\geq 0} \begin{aligned}[t]& \alpha ||\mathbf{S-UP^{\text{T}}}||_{F}^{2}+ \beta ||\mathbf{N-UVU^{\text{T}}}||_{F}^{2}\\+&
\gamma(||\mathbf{U}||_{F}^{2}+||\mathbf{V}||_{F}^{2}+||\mathbf{P}||_{F}^{2})
\end{aligned}
\label{eqn:model}\end{align}\normalsize
Here $\mathbf{U}$ contains the latent foci membership of the asker and his connections obtained by integrating network and content information, $\mathbf{P}$ shows the latent foci memberships of the words and $\mathbf{V}$ represents the correlation between the latent foci. $||\mathbf{U}||_{F}^{2}$, $||\mathbf{V}||_{F}^{2}$, and $||\mathbf{P}||_{F}^{2}$ are regularization terms introduced to prevent overfitting and $\gamma$ is the positive parameter for control the proportions of the regularization terms. The constraints $\mathbf{U}\geq 0$, $\mathbf{V} \geq 0$, and $\mathbf{P}\geq 0$ denote that the matrices do not contain negative elements. $\alpha$ and $\beta$ are positive parameters to control the effects of content and network proportions respectively.

We draw from the concepts of the social foci theory illustrated in Fig. \ref{fig:socfoctheo} to propose that users sharing a large amount of foci memberships with the asker in the question domain can effectively answer social media questions. The shared foci memberships of the asker with his social connections is given by the overlap between their corresponding rows in $\mathbf{U}$. The question domain in the latent foci space is obtained by combining the rows of $\mathbf{P}$ corresponding to the words in the question. Before formalizing these notions, we optimally derive the latent matrices $\mathbf{U}$, $\mathbf{V}$ and $\mathbf{P}$ by solving Eq. (\ref{eqn:model}).

\subsection{Deriving the Optimal Latent Matrices}
\label{subsec:optimal}

The problem presented in Eq. (\ref{eqn:model}) belongs to a class of constrained convex minimization problems. Motivated by \cite{ding2006orthogonal}, we describe an algorithm to find optimal solutions for $\mathbf{U}$, $\mathbf{V}$ and $\mathbf{P}$. The key idea is to optimize the objective with respect to one variable while fixing others. The three variables are iteratively updated until convergence.

From Eq.(\ref{eqn:model}), we let
\small\begin{align}
\mathcal{J}&=\begin{aligned}[t]
&\alpha||\mathbf{S-UP^{\text{T}}}||_{F}^{2}+\beta||\mathbf{N-UVU^{\text{T}}}||_{F}^{2}+\\
&\gamma(||\mathbf{U}||_{F}^{2}+||\mathbf{V}||_{F}^{2}+||\mathbf{P}||_{F}^{2})
\end{aligned}
\end{align}\normalsize
We then take the Lagrangian of the objective function $\mathcal{J}$. Let the Lagrange multiplier for the constraints $\mathbf{U} \geq 0$, $\mathbf{V} \geq 0$, and $\mathbf{P} \geq 0$  be $\Lambda_{\text{u}}$, $\Lambda_{\text{v}}$, and $\Lambda_{\text{p}}$ respectively. Then
\small\begin{equation}
\mathcal{L}=\mathcal{J}+tr(\Lambda_{\text{u}}\mathbf{U}^{\text{T}})+tr(\Lambda_{\text{v}}\mathbf{V}^{\text{T}})+tr(\Lambda_{\text{p}}\mathbf{P}^{\text{T}})
\end{equation}\normalsize
We compute the partial derivatives of the lagrangian $\mathcal{L}$ with respect to $\mathbf{U}$, $\mathbf{V}$, and $\mathbf{P}$ keeping the other variables fixed as shown below.
\small\begin{align}
\frac{\partial \mathcal{L}}{\partial \mathbf{U}}&=\begin{aligned}[t]
&2(\alpha(-\mathbf{SP}+\mathbf{UP}^{\text{T}}\mathbf{P})+\beta(-\mathbf{N}^{\text{T}}\mathbf{UV}-\mathbf{NUV}^{\text{T}}
\\& +\mathbf{UVU}^{\text{T}}\mathbf{UV}^{\text{T}}+\mathbf{UV}^{\text{T}}\mathbf{U}^{\text{T}}\mathbf{UV})+\gamma \mathbf{U})+\Lambda_{\text{u}}
\end{aligned}\notag\\
\frac{\partial \mathcal{L}}{\partial \mathbf{V}}&=\begin{aligned}[t]
&2(\beta(-\mathbf{U}^{\text{T}}\mathbf{NU}+\mathbf{U}^{\text{T}}\mathbf{UVU}^{\text{T}}\mathbf{U})+\gamma\mathbf{V})+\Lambda_{\text{v}}
 \end{aligned} \notag\\
\frac{\partial \mathcal{L}}{\partial \mathbf{P}} &=\begin{aligned}[t]&2(\alpha(-\mathbf{S}^{\text{T}}\mathbf{U}+\mathbf{PU}^{\text{T}}\mathbf{U})+\gamma \mathbf{P}) + \Lambda_{\text{p}}.
 \end{aligned}
 \label{eqn:diff}
\end{align}\normalsize
Substituting the KKT complementary conditions in Eq. (\ref{eqn:diff}) and rearranging we get the following update rules for latent matrices $\mathbf{U}$, $\mathbf{V}$, and $\mathbf{P}$.
\small \begin{align}
\mathbf{U}_{ij}&\leftarrow\begin{aligned}[t]
&\mathbf{U}_{ij}\sqrt{\frac{\alpha\mathbf{SP}+\beta(\mathbf{N}^{\text{T}}\mathbf{UV}+\mathbf{NUV}^{\text{T}})}
 {\alpha\mathbf{\mathbf{UP}^{\text{T}}\mathbf{P}}+\beta(\mathbf{UVU}^{\text{T}}\mathbf{UV}^{\text{T}}+\mathbf{UV}^{\text{T}}\mathbf{U}^{\text{T}}\mathbf{UV})+\gamma \mathbf{U}}}
\end{aligned}\notag\\
\mathbf{V}_{ij}&\leftarrow\begin{aligned}[t]
&\mathbf{V}_{ij}\sqrt{\frac{\beta\mathbf{U}^{\text{T}}\mathbf{NU}}{\beta(\mathbf{U}^{\text{T}}\mathbf{UVU}^{\text{T}}\mathbf{U})+\gamma \mathbf{V}}}
 \end{aligned} \notag\\
 \mathbf{P}_{ij}&\leftarrow\begin{aligned}[t]&\mathbf{P}_{ij}\sqrt{\frac{\alpha\mathbf{S}^{\text{T}}\mathbf{U}}{\alpha\mathbf{PU}^{\text{T}}\mathbf{U}+\gamma \mathbf{P}.}}
 \end{aligned}
\end{align}\normalsize

The optimization algorithm is summarized in Steps 1-7 in \textbf{Algorithm 1}. The square root on the update rules is added to ensure convergence \cite{ding2008nonnegative}. The correctness and convergence of the rules can be proved by the axillary function method \cite{lee2000algorithms}.

\subsection{Identifying Answerers from Foci Information}
\label{subsec:retpeop}
We now identify relevant answerers from the social connections of the asker using the latent matrices $\mathbf{U}$,$\mathbf{V}$ and $\mathbf{P}$. We first extract the words from the question attribute vector $\mathbf{w}_\text{q}$ and obtain the foci memberships of each word from the corresponding rows in matrix $\mathbf{P}$. We then compute the domain of the question in the latent foci space as a combination of individual word membership vectors as
\vspace*{-0.05in}
\small\begin{equation}
\vspace*{-0.1in}
\mathbf{d}_\text{q}=\sum_{\text{w}_\text{i}\epsilon \mathbf{w}_\text{q}}{\mathbf{P}_i},
\end{equation}\normalsize
where $\mathbf{d}_\text{q}$ represents the domain of the question q in the latent foci space and $\text{w}_\text{i}$ is the word corresponding to the $\text{i}^{\text{th}}$ row of $\mathbf{P}$.

We next compute the foci memberships of the asker and his social connections in the question domain. The Hadamard product of two vectors is the pointwise product of their respective elements, and it exactly captures this notion. For each question, we compute the Hadamard product of the row of $\mathbf{U}$ corresponding to the asker, $\mathbf{U}_{\text{A}}$ and the vector representing the question domain $\mathbf{d}_\text{q}$.
\small\begin{equation}
\mathbf{g}_\text{A}= \mathbf{U}_\text{A} \circ \mathbf{d}_\text{q},
\end{equation}\normalsize
where $\mathbf{g}_\text{A}$ contains the foci membership of the asker in the domain of the question. Similarly, we compute the foci memberships of each social connection of the asker in the domain of the question q by
\vspace*{-0.05in}
\small \begin{equation}
 \mathbf{g}_{\text{f}_\text{m}}= \mathbf{U}_{\text{f}_\text{m}} \circ \mathbf{d}_\text{q},
\end{equation}\normalsize
where $\text{f}_\text{m}$ is the $\text{m}^{\text{th}}$ social connection of the asker, $\mathbf{U}_{\text{f}_\text{m}}$ is the row of matrix $\mathbf{U}$ corresponding to $\text{f}_\text{m}$ and $\mathbf{g}_{\text{f}_\text{m}}$ contains the foci membership of $\text{f}_\text{m}$ w.r.t the domain of the question.

Finally, we find the overlap in foci memberships of the asker and his social connections in the question domain as
\small\begin{equation}
\mathbf{rs}{(\text{q},\text{A},\text{f}_{\text{m}})}=sim(\mathbf{g}_{\text{A}},\mathbf{g}_{\text{f}_\text{m}}),
\label{eqn:score}
\end{equation}\normalsize
where $\mathbf{rs}{(\text{q},\text{A},\text{f}_{\text{m}})}$ denotes the score of the answerer $\text{f}_{\text{m}}$ to the question q by the asker A. We sort the answerers according to their score and return them to the asker as a ranked list, $\mathbf{ra}$. Results with different similarity metrics is presented in Table \ref{tab:results}. The method for identifying answerers from foci information is summarized in Steps 8-11 in \textbf{Algorithm 1}. The quantity $\mathbf{rs}{(\text{q},\text{A},\text{f}_{\text{m}})}$ signifies the context in terms of network and content shared between asker A and his social connection $\text{f}_{\text{m}}$  in the domain of question q.

\small\begin{table}
\begin{center}

    \begin{tabular}{l}

    \toprule
     \textbf{Algorithm 1}: Automatic Identification of \\ Answerers to Social Media Questions\\
    \midrule
    \ \textbf{Input:} Question q of asker (A), friends and followers \\ of A, $\mathbf{f}_\text{A} = [\text{f}_{1},\text{f}_{2},....,\text{f}_{\text{m}}]$, Egonetwork of the asker (\textbf{N}), \\  user-word matrix of the asker and his connections (\textbf{S})\\   and $\left\{\alpha,\beta,\gamma,k\right\}$\\
    \ \textbf{Output :} A ranked list of the potential answerers $\mathbf{ra}$ \\

    \ 1: Initialize $\mathbf{U}$, $\mathbf{V}$, $\mathbf{P}$ randomly\\
    \ 2: \textbf{while} not convergent \textbf{do}\\
    \ 3: update \\
    \ 4: \small  $\mathbf{U}_{ij}\leftarrow \mathbf{U}_{ij} \sqrt{\frac{\alpha\mathbf{SP}+\beta(\mathbf{N}^{\text{T}}\mathbf{UV}+\mathbf{NUV}^{\text{T}})}
 {\alpha\mathbf{\mathbf{UP}^{\text{T}}\mathbf{P}}+\beta(\mathbf{UVU}^{\text{T}}\mathbf{UV}^{\text{T}}+\mathbf{UV}^{\text{T}}\mathbf{U}^{\text{T}}\mathbf{UV})+\gamma \mathbf{U}}}$ \\[2ex]
    \ 5: $\mathbf{V}_{ij} \leftarrow \mathbf{V}_{ij}\sqrt{\frac{\beta\mathbf{U}^{\text{T}}\mathbf{NU}}{\beta(\mathbf{U}^{\text{T}}\mathbf{UVU}^{\text{T}}\mathbf{U})+\gamma \mathbf{V}}}$ \\[2ex]
\ 6: $\mathbf{P}_{ij} \leftarrow  \mathbf{P}_{ij}\sqrt{\frac{\alpha\mathbf{S}^{\text{T}}\mathbf{U}}{\alpha\mathbf{PU}^{\text{T}}\mathbf{U}+\gamma \mathbf{P}}}$ \normalsize \\
\ 7: \textbf{end while}\\
\ 8: \small $\mathbf{w}_\text{q}=[\text{w}_{\text{q1}},\text{w}_{\text{q2}},...,\text{w}_{\text{ql}}]$, $\mathbf{d}_\text{q}=\sum_{\text{w}_\text{i}\epsilon \mathbf{w}_\text{q}}{\mathbf{P}_i}$ \\
\ 9: $\mathbf{g}_{\text{A}}= \mathbf{U}_{\text{A}} \circ \mathbf{d}_{\text{q}}$, $\mathbf{g}_{\text{f}_\text{m}}= \mathbf{U}_{\text{f}_\text{m}} \circ \mathbf{d}_{\text{q}}$\\
\ 10: $\mathbf{rs}{(\text{q},\text{A},\text{f}_{\text{m}})}=sim(\mathbf{g}_{\text{A}},\mathbf{g}_{\text{f}_\text{m}})$\\
\ 11: $\mathbf{ra}$=sort($\mathbf{rs}$) \normalsize \\
    \bottomrule
    \end{tabular}
    \end{center}
      \label{tab:Algorithm}
      \vspace*{-0.2in}
   \end{table}\normalsize
\subsection{Time Complexity}

The highest time cost results from updating the latent matrices in steps 4-6. In the updating terms, the complexity of the terms $\mathbf{SP}$ and $\mathbf{S}^\text{T}\mathbf{U}$ is low due to sparsity of $\mathbf{S}$. The terms $\mathbf{N}^\text{T}\mathbf{UV}$, $\mathbf{NUV}^\text{T}$ and $\mathbf{U}^\text{T}\mathbf{NU}$ have a complexity of $\text{O}(\text{mk}^2)$ where m is the number of friends and k is the number of latent dimensions due to the sparsity of $\mathbf{N}$. The terms $(\mathbf{U}(\mathbf{V}(\mathbf{U}^\text{T}\mathbf{U})\mathbf{V}^\text{T})$, $(\mathbf{U(V}^\text{T}\mathbf{(U}^\text{T}\mathbf{U)V})$ and $((\mathbf{U}^\text{T}\mathbf{U)V(U}^\text{T}\mathbf{U)})$  has a complexity of $\text{O}(\text{mk}^2)$  when computed as shown in the brackets. The complexity of $\mathbf{PU}^\text{T}\mathbf{U}$ and $\mathbf{UP}^\text{T}\mathbf{P}$ is $\text{O}(\text{(w+m)k}^2)$ where w is the number of words. Therefore, the overall complexity of a single iteration is $\text{O}(\text{(w+m)k}^2)$, which is low owing to the few number of latent dimensions. In addition, notice that steps 1-7 can be computed offline and only steps 8-10 are computed when the question is asked, further reducing the time required to identify answerers for a given question.

\section{Experiments}
\label{sec:dataana}
In this section, we first present a dataset of questions posted on Twitter and then conduct experiments to answer the following questions that help in understanding the framework better: How does the proposed framework perform in comparison to existing baselines? What is the effect of the amount of network and content information on the performance of the framework?
\subsection{Dataset}
\label{subsec:dataset}
The dataset consists of subjective questions from the social media platform Twitter. We follow the literature on questions in Twitter \cite{morris2010people} to construct a keyword set related to subjective questions. We append ``?'' to each keyword to collect questions from the Twitter Streaming API. Texts having ``?'' in online content are shown to be questions with high precision \cite{cong2008finding}. We deem replies to have been accepted by the asker if he has marked it as ``favorite'' or acknowledged the answerer by using ``thanks'' or ``thank you''. We mark the users who provided these answers as the ground truth for each question following \cite{hecht2012searchbuddies}. Some important statistics of the dataset are given in Table \ref{tab:dataset}. The first question was posted on Dec 27, 2013 and the last one on Jan 15, 2014. We use the methods in the public Twitter API to collect the friends, followers and public status messages of the asker to obtain the asker's social connections and their interests \cite{TwitterDataAnalytics2013}. We use the data to construct the ego network $\mathbf{N}$ and user-word matrix $\mathbf{S}$ for each asker.

\begin{table}
\begin{center}
    \begin{tabular}{l r}
    \toprule
     \textbf{Parameter} & \textbf{Statistics} \\
    \midrule
    \# of Questions & 1065 \\
    \# of Askers & 1026 \\
    \# of Selected Answers & 1450 \\
    \# of Followers and Friends of the askers & 966,117 \\
      Median \# of Followers and Friends per asker & 588 \\
      Median \# Tweets per user & 479\\
     \bottomrule
    \end{tabular}
    \vspace*{-0.2in}
    \end{center}
     \caption{Dataset containing questions posted in Twitter with statistics related to network and content information.}
  \label{tab:dataset}
  \vspace*{-0.2in}
  \end{table}

\subsection{Experiment Settings}
\label{subsec:evamet}
We introduce the following metrics to evaluate the performance of our framework: The Mean Reciprocal Rank (MRR) \cite{radev2002evaluating} is a measure of the overall likelihood of the framework to identify an answerer for a question, the Mean Average of Precision (MAP) \cite{bian2008finding} measures the potential satisfaction of the asker with the top K results and the Normalised Discounted Cumulative Gain (NDCG)@K considers the order within the top K rankings \cite{wang2013theoretical}.We use the following baselines to evaluate the performance of our framework.

\textbf{Random}: We randomly order the friends and followers of the asker 100 times and return the mean ordering.

\textbf{Aardvark} \cite{horowitz2010anatomy}: This paper describes a search engine which directed questions posted on the system to users with formulation to compute affinity with the asker and interest in the question topics. It does not consider the network structure and also does not contain experimental evaluations of its formulation.

\textbf{Content based Methods} \cite{riahi2012finding}: The paper focuses on community Q\&A like Yahoo! Answers and compares the similarity of the question topic with the interests of the answerers derived only from the their content. The interests were inferred by two topic models: LDA and the Segmented Topic Model (STM) \cite{du2010segmented}.

\textbf{Topic Sensitive Page Rank} \cite{zhou2012topic}: This paper employs a PageRank based approach to find subject matter experts in the question topic by combining network and content information of the potential answerers. The paper identifies topical authorities not considering the shared context between the asker and the answerers.

\textbf{Shared Foci}: This baseline measures the effect of shared user context. It computes the shared foci memberships of the asker and his social connections derived from either network ($\alpha$=0) or content ($\beta$=0) information. The question information is not taken into consideration. This also helps in evaluating methods using only network structure.

For initial experiments, we set the parameters in Eq. (\ref{eqn:model}) as follows. The regularization parameter is set at $\gamma$=0.01. The number of topics in the baselines and the number of foci k is set as 50. For initial evaluation of the framework, we choose $\alpha$=1 and $\beta$=1. The performance for different values of $\alpha$ and $\beta$ will be presented in future subsections.

\begin{table}
\begin{center}
    \begin{tabular}{l r r r}
    \toprule
     \textbf{Method} & \textbf{MRR} & \textbf{MAP@5} & \textbf{NDCG@5} \\
    \midrule
     Random & 1.20\% &1.12\% & 0.25\%\\
     Content-LDA & 1.56\% & 1.46\% & 0.30\%\\
     Content-STM & 1.93\% & 2.27\% & 0.50\%\\
     TSPR & 1.64\% & 1.63\% & 0.45\% \\
     Aardvark & 2.11\% & 2.53\% & 0.50\%\\
     Shared Foci (Network)&3.43\%&3.66\%&0.97\%\\
     Shared Foci (Content)& 3.60\%&3.87\%&1.17\%\\
     Our Model (Cosine) & 3.91\% & 4.63\% & 1.25\%\\
     Our Model (PCC) &   3.80\% &4.73\% & 1.31\%\\
     Our Model (Euclidean) & 4.36\% &5.54\% & 1.41\%\\
    \bottomrule
    \end{tabular}
    \vspace*{-0.2in}
    \end{center}
     \caption{Comparison of performance of the proposed framework with baselines.}
  \label{tab:results}
  \vspace*{-0.2in}
  \end{table}
\subsection{Performance Evaluation}
\label{subsec:perform}

The results of the evaluations are presented in Table \ref{tab:results}. From Table \ref{tab:results}, we can see that the proposed framework has outperformed the baselines by a considerable margin. We conducted a paired t-test to compare the performance of our framework with that of the baselines, and the results indicated the difference between them is significant. We make the following observations from the table.

The proposed framework gives more than 300\% improvement over random selection. We can see that simple formulations like the one in Aardvark that considers social network information performs on par with complex topical models using only content such as STM. The proposed framework also performs significantly better than methods identifying subject matter experts as answerers such as TSPR. This emphasizes the importance of social context to identify answerers to social media questions.

Considering shared foci between the asker and the answerer improves the performance over methods like Aardvark not utilizing community memberships. This shows the effectiveness of using social foci to exploit social context. Incorporating question information to consider the overlap only in the foci related to the question gives further improvement in the performance.

In summary, by designing approaches based on shared social context and exploiting the structure of social ties, the proposed framework can effectively identify answerers for social media questions in the dataset. Next, we wish to understand the effect of content and network information on the performance of our framework

\subsection{Effect of Content and Network Information}

\label{subsec:netcon}

\begin{figure}
\begin{center}

\includegraphics[scale=0.20]{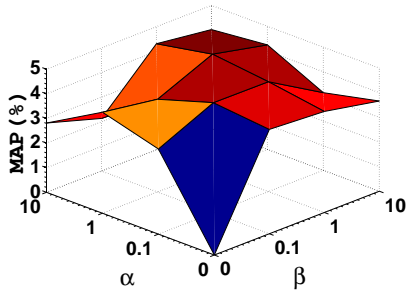}
\vspace*{-0.2in}
\end{center}
\caption{Effect of variation of content and network proportions on the framework performance for MAP.}
\label{fig:paraest}
\vspace*{-0.2in}
\end{figure}

In the model presented in Eq. (\ref{eqn:model}), $\alpha$ and $\beta$ control the proportion of the network and content information respectively. In order to evaluate the framework for different proportions of content and network, we set $\alpha=[0.1,1,10]$ and $\beta=[0,0.1,1,10]$ and plot the values for MAP in Fig. \ref{fig:paraest} arbitrarily using cosine similarity as the similarity metric. We make the following observations from the figure.

A general trend in Fig. \ref{fig:paraest} is a peak at the main diagonal of the $\alpha$ and $\beta$ axes and an off-diagonal dip. This shows that the framework works best for nearly equal proportions of network and content information. The MAP value is greater than 3\% for all $\alpha$ and $\beta$ except for low proportions of network information ($\alpha=10$, $\beta=[0, 0.1]$). This emphasizes the importance of social connections of the asker for identifying answerers to social media questions. The lowest performance across all parameter values is more than twice than random ordering indicating the effectiveness of the framework for low relative proportions of content or network information. Overall, the MAP value is above 3\% for different combinations of $\alpha$ and $\beta$ indicating the effectiveness of the framework for a wide range of parameter values.

In summary, the framework performs well over different proportions of network and content and is robust to their variation. An appropriate combination of network and content information can optimize the effectiveness of the framework for identifying answerers to social media questions.

 \subsection{Performance across Question Categories}
Literature on social media questions has identified kinds of questions people ask on Twitter. Recommendation, opinions,  factual and rhetorical questions are popular questions asked on Twitter \cite{morris2010people,paul2011twitter}. We select four categories related to subjective questions, ``Suggestions'', ``Opinion'', ``Favor'', and ``Rhetorical'', and evaluate our framework in identifying answerers for different question categories.

We employed human labelling to assign category labels to questions. Three people independently labeled the questions, and the labels were assigned using majority selection.  Employing this procedure, 93.5\% of the questions were assigned to either of the four categories and the framework was evaluated on them. The results of the evaluations are presented in Table \ref{tab:casestudy}. The distribution of different question categories is given in the first column. The performance for different categories is listed in the other columns. The improvement over \cite{horowitz2010anatomy}, the nearest baseline not a part of our method, for different question categories is shown in the brackets.

From the table, we see that the framework gives considerable improvements over all the selected question categories. A paired t-test suggested that the improvements are significant indicating that the framework is effective in finding answerers to a wide range of question categories in Twitter. The best performance can be seen in ``Suggestions'' and ``Favor'' categories and the performance in ``Opinions'' is relatively lower. These results suggest that identifying answerers for the ``Opinion'' category might depend on additional factors such as similarity of views in a given topic. The framework gives the lowest performance for questions in the ``Rhetorical'' category. Rhetorical questions are classified as conversational questions in the literature \cite{harper2009facts}. They might be used as an expression of opinion or to initiate a conversation and not to express an information need.

\begin{table}
\begin{center}
    \begin{tabular}{l l l l}
    \toprule
     \textbf{Categories} & \textbf{Parts} & \textbf{MRR}&\textbf{MAP@5} \\
     \midrule
  Suggestions& 39.83\%&4.27\%(+2.23\%)&4.68\%(+1.78\%) \\
  Opinion& 16.42\%& 2.67\%(+1.43\%)&2.38\%(+1.61\%)\\
Favor& 30.51\% &3.65\%(+1.55\%)&4.39\%(+1.01\%) \\
Rhetorical& 6.74\% &1.75\%(+1.17\%)& 0\%(+0\%) \\
       \bottomrule
    \end{tabular}
    \vspace*{-0.2in}
    \end{center}
    \caption{Performance for different question categories.}
  \label{tab:casestudy}
  \vspace*{-0.2in}
  \end{table}

\section{Conclusion and Future Work}
\label{sec:conclusion}

Online social media provides a new platform for people seeking information from their social connections. Social media questions represent a form of information seeking behavior of users. Questions are subjective and personal to the asker, and his social context is useful to identify answerers. We draw from sociological theories to present a novel framework to infer the shared context between the asker and the answerers in the question domain. We evaluate the framework on questions on Twitter and demonstrate its effectiveness in identifying answerers. The framework is robust to a wide range of proportions of network and content information and categories of social media questions. The paper provides the first framework with experimental evaluations to identify answerers to questions in social media.

Frameworks exist to identify answerers to factual questions prevalent in community Q\&A platforms like Yahoo Answers and StackOverflow. Incorporating concepts from them will enable us to tackle more diverse questions. During situations like natural disasters, social media users propagate requests for help throughout the network. Identifying answerers in these situations will require an understanding of information propagation and information seeking behavior. Identifying users providing misinformation to questions in social media will help to increase the effectiveness of social media as a quality information source.

\section{Acknowledgements}
This material is based upon work supported by, or in part by, Office of Naval Research (ONR) under grant number N000141010091.

\bibliographystyle{aaai}
\bibliography{sigproc}
\end{document}